# A Theoretical Approach for Dynamic Modelling of Sustainable Development

CORINA-MARIA ENE, ANDA GHEORGHIU, ANCA GHEORGHIU
Hyperion University, ROMANIA
corina.maria.ene@gmail.com, andagheorghiu@yahoo.com, anca.gheorghiu@gmail.com
www.hyperionline.ro

*Abstract: This article presents a theoretical model for a dynamic system based on sustainable development. Due to the relatively absence of theoretical studies and practical issues in the area of sustainable development, Romania aspires to the principles of sustainable development. Based on the concept as a process in which economic, social, political and natural environment are combined in order to sustain planet management, our goal is to promote an economic tool for Romanian decision-makers in order to evaluate scenarios and planning options.*

*Key-Words*: sustainable development, economic system, economic development, economic welfare, resources scarcity, natural environment

## 1. Introduction

"Bruntland Report" defines sustainable development as "…a process of change in order to create a harmony environment between resource exploitation, investments, technological and institutional changes to enhance the current and the future potential of human needs".

Starting from this statement, it is obvious that the concept of sustainable development tries to test the coexistence of environmental protection and economic development in a global and long – term point of view.

## 2. Conceptual framework on sustainable development

Sustainable development is neither a doctrine nor a theory, much less a synthesis between economics and ecology. It is a pragmatic approach to implement economic tools for planet management. Sustainable development is a new term for an old idea: there is no viable economy without natural resources and no resources management without economic rationale. We can distinguish two type of redundant views related to the concept of sustainable development: a vision of a global economic space and an environmental vision.

Global economic vision for sustainable development involves complex conditions and factors that enable the revenues to increase involving economic welfare aspects as population growth rule, classification of resources reflecting their relative scarcity, changing production and consumption structure in order to maintain the stock of scarce resources.

The environmental vision of sustainable development involves the management and the maintenance of resource stocks and factors yielding a consistent productivity, at least, in the spirit of equity between generations and countries. The resource stock includes two different components: "artificial" capital stock (includes all production factors made by human civilisation) and "natural" capital meaning all renewable and non-renewable resources (water, fauna, flora and soil).

Issues on sustainable development include the following: (1) resizing of economic growth given a more equitable resources distribution and an increased production quality, (2) conditions to eliminate poverty in the context of people's essential needs assurances such as employment, food, energy, water, housing and health, (3) reduction of uncontrolled population growth, (4) preserving and enhancing of natural resources, (5) maintenance of ecosystems' diversity, (6) monitoring the environmental impact on economic development, (7) technology diversion and risks control, (8) government decentralization, (9) increasing people's participation of decisions on environmental and economic problems.

## 3. Methodology. Systems and Components

In order to create a sustainable development model we are forced to use some system dynamics techniques to formulate and simulate such a model. To simplify real world phenomena and interrelationship of various variables we turn into some abstractions.

First of all, it is necessary to identify the system we want to model and its components. It is important to consider all components and to prioritize them. Usually, all indicators are qualitatively and/or quantitatively measured. It is our choice to choose them taking care of sustainable development system behaviour concerning human, economic and natural components.

The human component concerns the social dimension of sustainable development. There is no sustainable development without a strong human basis centred on individual development (personal education, abilities, opportunities etc.). The social sector (education system, health, social insurances etc.) is another facet of the same problem, as well as government sector (laws, rules).

The economic component implies all economic aspects on sustainable development (public economics, economic growth, welfare, life quality and infrastructure).

The natural component represents the environmental aspect of sustainable development (natural resources and environment).

Now, it is obvious that a sustainable development model must integrate the social, economic and environmental components. So, the conceptual model design could be as follows (Figure 1). The next step is to segment its components and to measure indicators (Figure 2).

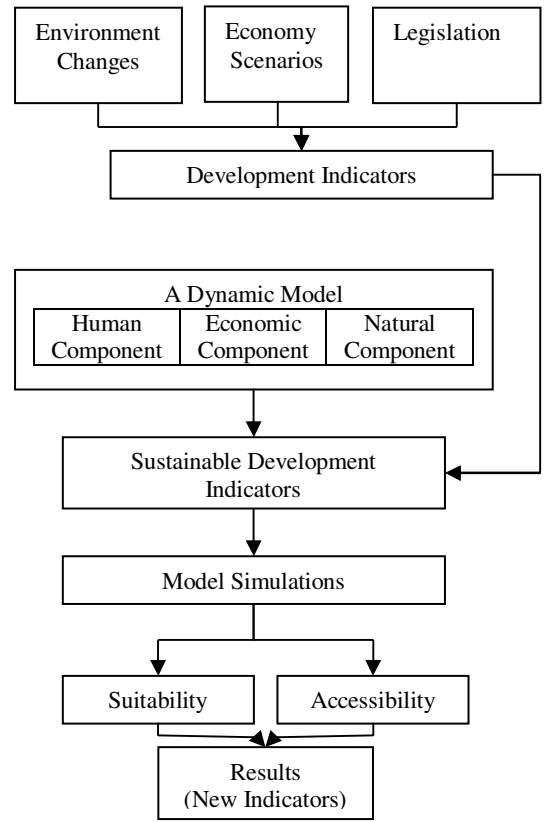

Figure 1 - The conceptual sustainable development model design

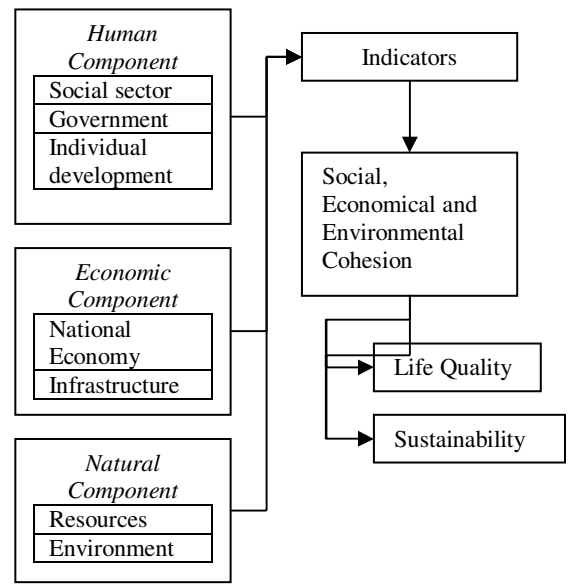

Figure 2 - Segmented components of the conceptual sustainable development model

Economic component concerns national economy and infrastructure. Natural component refers to natural resources and environment.

First of all, social sector implies population and it is focused on social security, health care, social organizations, domestic income composition etc. It is important to measure population immigration and emigration referring to births (in-migration) and deaths (out-migration). Both are modelled based on a logarithmic function of form (Mayerthaler, 2009):

$$PPFL = e^{\alpha_0 + \alpha_1 POP + \alpha_3 L + \alpha_4 GL} \quad (1)$$

PPFL – population flow;
$\alpha_1, \alpha_2, \alpha_3, \alpha_4$ – parameters;
POP – population number;
L – national land;
GL – national green land.

Government sector is represented by public finances, taxes, national gross domestic product per capita (GDPP) etc.

$$GDPP = [C+I+G + (X - M)]/POP, \quad (2)$$

GDP is a sum of Consumption (C), Investment (I), Government Spending (G) and Net Exports (X - M); POP – population.

Individual development refers to elements of education, recreation, leisure, social integration, standard of living etc. To measure this component we use human development index (HDI), education index (EI), health index (HI) and income index (II).

HDI has three dimensions, measured by one or two indicators each:
- leading a long and healthy life - life expectancy at birth (LE);
- education - adult literacy rate, gross primary, secondary and tertiary enrolment (EI);
- a decent standard of living - GDP per capita (GDPI).

$$HDI = \frac{GDPI + EI + LE}{3} \quad (3)$$

HDI can be replaced by Human Poverty Index (HPI).

$$HPI = \left[\frac{1}{3}\left(P_1^\alpha + P_2^\alpha + P_3^\alpha\right)\right]^{1/\alpha} \quad (4)$$

$P_1$ – probability of not surviving to age 40 (times 100);
$P_2$ – adult illiteracy rate;
$P_3$ – average of people without access to safe water and children underweight.

As $\alpha$ rises greater weight is given to the dimension in which there is most deprivation.

$\alpha = 1$ implies simple average (perfect substitutability);

$\alpha = \infty$, HPI = highest value (no substitutability);

$\alpha = 3$ gives additional, but not overwhelming weight to areas of most acute deprivation.

The economic component consists of two elements: national economy (economic system) and infrastructure.

The infrastructure implies economic supply system (energy, water, and food), tourism objectives, and facilities for education, health, research and development. Increasing population generates increasing demand for infrastructure capacities. Economic infrastructures are key assets that are needed to support the long term growth of the economy. In order to measure economic infrastructure we use The Macquarie Global Infrastructure Index. It includes the following indices: Oil & Gas Index, Telecom Equipment Index, Transport Services Index, Utilities Index, Electricity Index, Gas Distribution Index, Multi-utilities Index, Water Index and much more.

The national economy model simulates industrial production, agricultural production, consumption, commerce and trade processes, labour and employment. The production modelling is based on a modified Cobb-Douglas production function with adapted parameters.

$$Y = A \cdot L^\alpha \cdot K^\beta \cdot P^{1-\alpha-\beta} \quad (5)$$

Y is production output (industrial and agricultural production);
A is total factors productivity;
L is employment in industrial and agricultural sector;
K is industrial and agricultural capital input;
P is land input;
$\alpha$ is employment elasticity and $\beta$ is capital elasticity.

Currently the parameters of industrial and agricultural capital and employment elasticity are constants, whose values can be obtained from a regression analysis.

Natural component is represented by resources and environment (atmosphere and hydrosphere, all natural resources, ecosystems, material recycling and pollution).

The total land area consists of different land types such as forest, agricultural land, unused land, land used for urban & industrial settlements etc. A transition between different land types is driven by economical and environmental factors. We can use an Index of this expansion/increment of the agricultural and urban areas.

## 4. Model Indicators

These indicators must aggregate all three dimensions elements we previously discussed (Figure 3).

The life quality could be calculated using the following equation (similar to Zgurovsky, 2009):

$$I_{lq} = \sqrt{I_s + I_{ec} + I_n} \cdot \cos\alpha \quad (6)$$

$I_{lq}$ – population life quality;
$I_s$ – index of social component;
$I_{ec}$ – index of economic component;
$I_n$ – index of environmental balance (ecological dimension);

$$\cos\alpha = \frac{I_s + I_{ec} + I_n}{\sqrt{3(I_s^2 + I_{ec}^2 + I_n^2)}}$$ represents the degree of component harmonization.

As it is shown by Figure 3, life quality gives the measures of sustainable development. It could be calculated as a modulus of a complex number using the following equation (Zgurovsky, 2009):

$$I_{sd} = |jI_{\sec} + I_{lq}| = \sqrt{I_{\sec}^2 + I_{lq}^2} \quad (7)$$

$I_{sd}$ – index of sustainable development;
$I_{\sec}$ – index of population social security;
$I_{lq}$ – population life quality.

The model is relatively simple, but it is not fully completed. It must be calibrated and some simulations based on statistical dates are necessary.

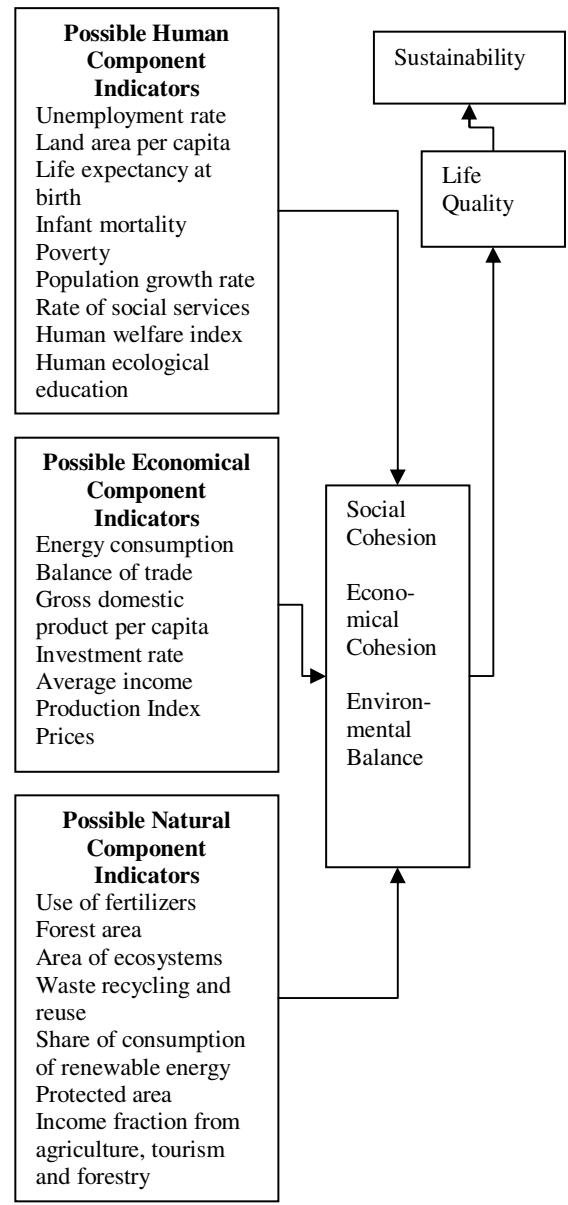

Figure 3 - Aggregate elements of sustainable development model

These indicators are an effective means of quantifying and measuring progress towards sustainable development. They help to create a "vision" of what sustainable development means in our everyday lives.

We recommend that the Governments brought forward a set of headline indicators with:
• Key priorities that better reflected sustainability principles and policy;

• A radically different approach to measuring economic progress;
• Challenging targets as to how they should move over time; and
• More effective machinery for acting on adverse trends.

## 5. Romania's Sustainable Development Strategy

Romania adopted a National Sustainable Development Strategy (2013-2020-2030) in 2008. It aims to connect Romania to a new philosophy of development, adopted by the European Union and widely shared globally—that of sustainable development.

This Strategy sets specific objectives for moving, within a reasonable and realistic timeframe, towards a new model of development that is capable of generating high value added, is motivated by interest in knowledge and innovation, and is aimed at continued improvement of the quality of life and human relationships in harmony with the natural environment.

In terms of general orientation, this paper addresses the following **strategic objectives** for the short, medium and long run:
- Horizon 2013: To incorporate the principles and practices of sustainable development in all the programs and public policies of Romania as an EU Member State.
- Horizon 2020: To reach the current average level of the EU countries for the main indicators of sustainable development.
- Horizon 2030: To get significantly close to the average performance of the EU Member States in that year in terms of sustainable development indicators.

The implementation of these strategic objectives will ensure high rates of economic growth in the medium and long run and, as a result, a significant reduction of social and economic disparities between Romania and the other Member States of the European Union. Considering the main indicator that measures convergence in real terms, Gross Domestic Product per capita adjusted for standard purchasing power parity (PPP), the implementation of the Strategy enables Romania to exceed in 2013 half of the current EU average, to approach 80% of the EU average in 2020 and to rise slightly above the EU average in 2030.

## 6. Conclusions

The goal of the proposed model is to explore alternative scenarios to improve the population life quality and sustainability in a national context. Our next step will be to realize a calibration based on data for Romania, including population, land use and energy production. Research work and development continues and the model framework will be improved. Once fully developed, calibrated and tested, the proposed theoretical framework will be an important tool in assisting Romanian decision makers in sustainable development planning, as well as an educational mean in understanding how various national systems function.